\documentstyle[aps,eqsecnum,preprint,tighten]{revtex}
\begin{document}
\draft
\preprint{\vbox{Submited to {\it Physical Review \bf{D}}\hfill
IFUSP/P-1171\\}}
\title{A QCD Sum Rule Approach to the $s\rightarrow d\gamma$
Contribution to the $\Omega^-\rightarrow \Xi^-\gamma$ Radiative Decay}
\author{M. Nielsen, L.A. Barreiro, C.O. Escobar and R. Rosenfeld%
\thanks{Address after October 1$^{st}$:
Instituto de F\'{\i}sica Te\'orica, Universidade Estadual Paulista,
Rua Pamplona 145, 01405-900 S\~ao Paulo-SP, Brazil}}
\address{Instituto de F\'{\i}sica, Universidade de S\~ao Paulo \\
Caixa Postal 66318 - 5389-970 S\~ao Paulo, S.P., Brazil}
\maketitle

\begin{abstract}
QCD sum rules are used to calculate the contribution
of short-distance single-quark transition $s\rightarrow d
\gamma$, to the amplitudes of the hyperon radiative decay,
$\Omega^-\rightarrow \Xi^-\gamma$.
We re-evaluate the Wilson coefficient of the effective operator
responsible for this transition.
 We obtain a branching ratio
which is comparable to the unitarity limit.
\end{abstract}
\newpage
\section{Introduction}
\label{sec:intro}
%
Hyperon radiative decays constitute an interesting class of processes
that has been studied intensively, both theoretically and experimentally,
for the past $20$ years. However, despite these efforts we still lack a global
understanding of these processes \cite{lachpolones}.

Some of these decays may be sensitive to the  $s \rightarrow d \gamma$
transition, which tests the Standard
Model at the one-loop level and is  similar to the
$b \rightarrow s \gamma$ transition recently  seen at Cornell which has
generated a large amount of interest \cite{bsg}.
Therefore, it would be extremely interesting to obtain
any experimental information on it. In this paper we re-examine the
assertion \cite{Singer} that the decay $\Omega^- \rightarrow \Xi^- \gamma $
can provide a window to the $s \rightarrow d \gamma$ transition.
In the following we briefly review the arguments that led to this conclusion.

There are two distinct contributions to hyperon radiative
decays, namely
short- and long-distance contributions. In short-distance processes, the
photon is emitted at a close distance (typically $x \simeq 1/M_W$) from the
weak interaction vertex whereas in long-distance processes
($x \simeq$ confinement radius) one can separate the weak from the
electromagnetic vertices.

At the quark-level, there are in general three types of contribution to
hyperon radiative decays, as shown in Fig. 1.
The first diagram (Fig.1a) corresponds to a W-exchange and is
a long-distance process,
whereas Fig. 1b (single-quark transition $s \rightarrow d \gamma$) and
Fig. 1c (penguin diagram) are short-distance processes.

It is known that the $s \rightarrow d \gamma$ transition cannot be the
dominant one in all hyperon decays \cite{GW}.
This is to be expected since  W-exchange is the
main term in the $\Delta S=1$ effective Hamiltonian.
However, there are two hyperon radiative decays, namely
$\Omega^- \rightarrow\Xi^- \gamma$ and $\Xi^- \rightarrow \Sigma^-
\gamma$, where the quark content
of the initial and final state baryons does not allow a
W-exchange contribution.
Therefore, only the  $s \rightarrow d \gamma$  and penguin diagrams
contribute at the quark level to the decay.

Let us concentrate first on  $\Xi^- \rightarrow \Sigma^- \gamma$.
It has been shown \cite{KE} that the penguin contribution for this
process is negligible. In Ref.\cite{Singer2}, the $s \rightarrow d \gamma$
contribution to the branching ratio was calculated to be
$BR^{s \rightarrow d \gamma} (\Xi^- \rightarrow \Sigma^- \gamma) =
1.8 \times 10^{-5}$.
However, although the W-exchange process does not contribute directly to
 $\Xi^- \rightarrow \Sigma^- \gamma$, it does contribute to
$\Xi^- \rightarrow \pi^- \Lambda$, where the final state can then
re-scatter
as $\pi^-  \Lambda \rightarrow \Sigma^- \gamma$. This long-distance
contribution
has been estimated as $BR^{LD}( \Xi^- \rightarrow \Sigma^- \gamma) =
1.7 \times 10^{-4}$ \cite{KS}, close to the experimental result
$BR^{Exp}( \Xi^- \rightarrow \Sigma^- \gamma) = (1.27 \pm 0.23)
\times 10^{-4} $ \cite{PDB}.

This leaves us with the process  $\Omega^- \rightarrow
\Xi^- \gamma$ as the only candidate for probing the
$s \rightarrow d \gamma$ transition in hyperon radiative decays.
Let us summarize what is known about this process so far.
There is an experimental upper limit determined recently \cite{ivone}:
\begin{equation}
BR^{Exp.}( \Omega^- \rightarrow \Xi^- \gamma) < 4.6  \times 10^{-4} .
\label{exp}
\end{equation}
This is well above the unitarity limit \cite{KS} :
\begin{equation}
BR^{Unitarity}( \Omega^- \rightarrow \Xi^- \gamma) > 0.8  \times 10^{-5} .
\label{unitarity}
\end{equation}
The penguin contribution has been estimated as \cite{KE} :
\begin{equation}
BR^{Penguin}( \Omega^- \rightarrow \Xi^- \gamma) \simeq 5
\times 10^{-6} .
\label{penguin}
\end{equation}
Kogan and Shifman \cite{KS} have also calculated the long-distance
contribution from the dominant
$\Xi^0 \pi^-$ intermediate state as :
\begin{equation}
BR^{LD}( \Omega^- \rightarrow \Xi^- \gamma) = (1 \mbox{--}  1.5)
\times 10^{-5}  .
\label{long}
\end{equation}
The contribution from the single-quark $s \rightarrow d \gamma$ transition
has  been re-evaluated recently \cite{ara} yielding :
\begin{equation}
BR^{s \rightarrow d \gamma} (\Omega^- \rightarrow \Xi^- \gamma) =
8.0 \times 10^{-7}
\label{arasin}
\end{equation}
This result corrects a previous calculation \cite{Singer}, where it
was claimed that the single
quark transition should be the dominant contribution to the $\Omega^-$
radiative decay.
Both calculations \cite{ara,Singer} were based on
estimating
the hadronic matrix element of the effective operator  describing the
single-quark transition, using SU(6) quark-model wave functions
and assuming unit overlap.

Due to the importance of this process, we believe that a more
reliable estimate of the single-quark $s \rightarrow d \gamma$ contribution
to the radiative decay $\Omega^- \rightarrow \Xi^- \gamma$ is
in order. This is the purpose of the present work, where we perform a
calculation of the same matrix element using the well known
technique of QCD sum rules \cite{SVZ2}.

The QCD sum rule approach
was successfully applied to the $\Sigma^+ \rightarrow p \gamma$ decay
by Balitisky and collaborators \cite{russos}, who calculated in this way
the dominant W-exchange. QCD sum-rules were also employed before for the
calculation of the single-quark contribution \cite{goldman} to
the $\Sigma^+$ and $\Xi^-$ radiative decays. Although the single quark
transition is not the dominant contribution in these decays, the
calculation exposed the role of non-perturbative
corrections which alter the naive picture of the single-quark transition
in a major way: it showed that it is not a good approximation
to assume a unity overlap for the quark-model wave functions.

This paper is organized as follows. Section II discusses the effective
operator
for the $s \rightarrow d \gamma$ transition. In Section III we derive the
sum rules and in Section IV we present the results and conclusions.

\section{The Effective Operator for $s \rightarrow d \gamma$ Transition}
\label{sec:operator}
We are interested in computing the contribution of the quark level
$s \rightarrow d \gamma$ transition to the radiative decay $\Omega
\rightarrow
\Xi \gamma$. This transition does not occur at tree-level in the Standard
Model. Hence, it could in principle provide a good test to the
Standard Model at the $1$-loop level, being sensitive also to new Physics.

The effective hamiltonian (with the heavy quarks $c,b$ and $t$ as well as
electroweak gauge bosons integrated out ) that describes $| \Delta S = 1 |$
transitions is given by :
\begin{equation}
{\cal H }_{\mbox{eff.}}^{| \Delta S = 1 | } = \frac{-4 G_F}{\sqrt{2}} \;
\lambda_u \; \sum_{k=1}^8 \; c_k(\mu) {\cal O}_k(\mu) \; ,
\label{H}
\end{equation}
where $G_F$ is the Fermi constant, and we use the notation $\lambda_i$ to
denote the following product of elements of the Cabibbo-Kobayashi-Maskawa
(CKM) matrix, $\lambda_i  =
V_{si} V^{\ast}_{id} $ . The Wilson coefficients $c_k$ can be computed
perturbatively and $\{ {\cal O}_k \}$ is a complete set of operators
written as:
\begin{eqnarray}
{\cal O}_1 &=& (\bar{u}_{\alpha} \gamma_{\mu} P_L s_{\beta} )
              (\bar{d}_{\beta} \gamma^{\mu} P_L u_{\alpha} )
\nonumber\\*[7.2pt]
{\cal O}_2 &=& (\bar{u}_{\alpha} \gamma_{\mu} P_L s_{\alpha} )
              (\bar{d}_{\beta} \gamma^{\mu} P_L u_{\beta} )
\nonumber\\*[7.2pt]
{\cal O}_3 &=& (\bar{d}_{\alpha} \gamma_{\mu} P_L s_{\alpha} )
\sum_{q=u,d,s} (\bar{q}_{\beta} \gamma^{\mu} P_L q_{\beta} )
\nonumber\\*[7.2pt]
{\cal O}_4 &=& (\bar{d}_{\alpha} \gamma_{\mu} P_L s_{\beta} )
\sum_{q=u,d,s} (\bar{q}_{\beta} \gamma^{\mu} P_L q_{\alpha} )
\nonumber\\*[7.2pt]
{\cal O}_5 &=& (\bar{d}_{\alpha} \gamma_{\mu} P_L s_{\alpha} )
\sum_{q=u,d,s} (\bar{q}_{\beta} \gamma^{\mu} P_R q_{\beta} )
\nonumber\\*[7.2pt]
{\cal O}_6 &=& (\bar{d}_{\alpha} \gamma_{\mu} P_L s_{\beta} )
\sum_{q=u,d,s} (\bar{q}_{\beta} \gamma^{\mu} P_R q_{\alpha} )
\nonumber\\*[7.2pt]
{\cal O}_7 &=& \frac{i e}{16 \pi^2} m_s (\bar{d}_{\alpha} \sigma_{\mu \nu} P_R
              s_{\alpha} ) F^{\mu \nu}
\nonumber\\*[7.2pt]
{\cal O}_8 &=& \frac{i g_s}{16 \pi^2} m_s (\bar{d}_{\alpha} \sigma_{\mu \nu}
 T_{\alpha \beta}^a  P_R
              s_{\beta} ) G^{a \mu \nu}  ,
\label{ope}
\end{eqnarray}
where $ P_{\tiny{\begin{array}{l}
L\\R\end{array}}}=\frac{1}{2}(1\mp\gamma_5) $ , $\sigma_{\mu \nu} =
 i [ \gamma_{\mu}, \gamma_{\nu} ] /2 $; $F^{\mu \nu}$ and $G^{a \mu \nu}$ are
the electromagnetic and color field tensors respectively.

We are interested in the Wilson coefficient $c_7(\mu = m_s)$, which controls
the short distance contribution to the process $s \rightarrow d \gamma$.

The Wilson coefficients are first computed at the scale $\mu = M_W$ in zeroth
order in QCD. QCD corrections are included by evolving these coefficients down
to $\mu = m_s$ using the renormalization group equations. This evolution
is accomplished in three steps. Schematically one has :
\begin{equation}
c_k(\mu = m_s) = U^3_{kl}(m_s,m_c) \; U^4_{ln}(m_c,m_b) \; U^5_{nm}(m_b,M_W) \;
                 c_m(M_W) .
\end{equation}
The evolution matrix with $f$ active quark flavors in the leading
logarithmic approximation can be written as \cite{buras}  :
\begin{equation}
U^f(m_1,m_2) = V T V^{-1} \; ,
\end{equation}
where $V$ is a matrix that diagonalizes the transpose of the  $8 \times 8$
anomalous dimension matrix $\gamma$ :
\begin{equation}
\gamma_D = V^{-1} \gamma V  \; ,
\end{equation}
and $T$ is a diagonal matrix with elements given by :
\begin{equation}
T_{ii} = \left[ \frac{\alpha_s(m_2)}{\alpha_s(m_1)} \right]^
                { (\gamma_D)_{ii}/2 \beta_0 }  \;  ,
\end{equation}
where $\beta_0 =11 - 2 f/3$ and in this approximation the running of the strong
coupling constant is described by :
\begin{equation}
\alpha_s(m_2) = \frac{\alpha_s(m_1)}{1 - \frac{\beta_0}{2 \pi} \alpha_s(m_1)
\ln(m_1/m_2) } .
\label{alpha}
\end{equation}

We now turn to a discussion of the initial conditions.
Due to our choice of factoring out $\lambda_u$ in eq.(\ref{H}), we have:
\begin{equation}
c_2(M_W) = 1  \;\;\; , \;\;\; c_{1,3-6}(M_W) = 0\; .
\label{initial}
\end{equation}
The values for $c_7$ and $c_8$ at $\mu = M_W$ are given in terms of functions
 arising from one-loop diagrams with an external photon
and a gluon respectively \cite{InamiLim,gustavo}. These function
$F_2(x_i)$ and $D(x_i)$, where $x_i = m_i^2/M_W^2$ and $m_i$ is the mass of
the quark running in the loop, are given by:
\begin{eqnarray}
F_2(x) &=&
Q \left[ \frac{x^3-5 x^2-2x}{4(x-1)^3} + \frac{3 x^2 \ln(x)}{2 (x-1)^4} \right]
+ \frac{2 x^3 + 5 x^2 - x}{4 (x-1)^3 } -  \nonumber \\
& &  \frac{3 x^3 \ln(x)}{2 (x-1)^4}   \; ;
\\
D(x) &=& \frac{x^3 - 5x^2 - 2x}{4(x-1)^3} + \frac{3x^2 \ln(x)}{2(x-1)^4} \; ,
\end{eqnarray}
where $Q$ is the charge of the internal quark.

In table 1 we show the contributions to $F_2$ from the different internal
quarks and its product with the relevant CKM matrix elements.
We use the following values for the quark masses and $\lambda_i$ :
$m_t = 175$ GeV, $m_c = 1.5$ GeV, $m_u = 5$ MeV, $\lambda_u = 0.21$,
$\lambda_c = -0.21$ and $\lambda_t = -(1.2 - 7.2) \times 10^{-4}$,
where the CKM elements were obtained from the standard parametrization of the
Particle Data Book neglecting CP violation, that is taking $\delta_{13} = 0$.

{}From this table we see that, as opposed to the $b \rightarrow s \gamma$ case,
the top quark contribution is not the dominant one for $s \rightarrow d \gamma$
and we must take into account
both $t-$ and $c-$ quarks in computing the initial values for $c_7$ and
$c_8$, which are given by :
\begin{eqnarray}
c_7(M_W) &=& \frac{-1}{2} \left( \frac{\lambda_t}{\lambda_u} F_2(x_t) +
                            \frac{\lambda_c}{\lambda_u} F_2(x_c) \right) ; \\
c_8(M_W) &=& \frac{-1}{2} \left( \frac{\lambda_t}{\lambda_u} D(x_t) +
                            \frac{\lambda_c}{\lambda_u} D(x_c) \right) .
\end{eqnarray}

In order to compute the evolution matrix we use the anomalous dimension matrix
given in ref. \cite{buras} and $\alpha_s(M_W) = 0.124$. However, from
eq.(\ref{alpha}) one would get a value for $\alpha_s(m_s) > 1$, since for
consistency with the leading-log anomalous dimension matrix we have to
compute the running of $\alpha_s$ in one-loop only.
In this case we use $\alpha_s(m_s) = 1$.
We find:
\begin{equation}
c_7(m_s) = -0.50 c_2(M_W) + 0.27 c_7(M_W) + 0.13 c_8(M_W) .
\label{result1}
\end{equation}

The numerical values for the initial conditions for $c_7$ and $c_8$ are:
\begin{equation}
c_7(M_W) = (2.1 - 7.7) \times 10^{-4}  \;\;\;\; , \;\;\;\;
c_8(M_W) = (1.4 - 4.2) \times 10^{-4} \; ,
\end{equation}
and from Eq.(\ref{result1}) we see that $c_7(m_s)$ is insensitive to the
the values of $c_7(M_W)$ and $c_8(M_W)$, being dominated by $c_2(M_W)$.
This result is in contrast to what happens in the $b \rightarrow s \gamma $
case. It can be understood since in the $ s \rightarrow d \gamma$ case we have
$c_2(M_W) \propto \lambda_u$ and $c_{7,8}(M_W)$ is effectively proportional
to $\lambda_t$ with
$\lambda_t \ll \lambda_u$; in the $b \rightarrow s \gamma$ case,
one has $c_2(M_W) \propto |V_{bc} V^{\ast}_{cs} | \simeq
|V_{bt} V^{\ast}_{ts} | \propto c_{7,8}(M_W) $ .

Unfortunately, Physics beyond the Standard Model would contribute mostly to
 $c_{7,8}(M_W)$ and from our calculation we see that unless the modification
implies in an increase of  $c_{7,8}(M_W)$ by roughly four orders of magnitude,
no significant contribution will be made to the $s \rightarrow d \gamma$
process.

Using equations (\ref{ope}), (\ref{initial}) and (\ref{result1}) we arrive
at the main result of this section:
\begin{equation}
{\cal H }_{\mbox{eff.}}^{s \rightarrow d \gamma } = \frac{-4 G_F}{\sqrt{2}} \;
\lambda_u \;  c_7(m_s) {\cal O}_7
= (0.28) \; \frac{i e G_F}{16 \pi^2} m_s (\bar{d}_{\alpha} \sigma_{\mu \nu} P_R
              s_{\alpha} ) F^{\mu \nu} \; .
\label{eq:eff}
\end{equation}

\section{QCD Sum Rules}
\label{sec:QCDSR}
The subject of this section is the evaluation of the
$s\rightarrow d\gamma$ contribution to the four multipole
amplitudes that appear in the transition matrix related to the
radiative decay $\Omega^-\rightarrow\Xi^-\gamma$, using a QCD sum
rule approach.
\subsection{Formalism}

We start with the three-point correlation function
\begin{equation}
\Pi_\mu(Q_i,Q_f) = -i\int d^4x d^4y e^{iQ_fy} e^{-iQ_ix}
\langle 0|{\rm T}[\eta^\Xi(y){\cal H}^{s \rightarrow d \gamma}_{eff}(0)
\overline{\eta}_\mu^\Omega(x)]
|0\rangle \; ,
\label{correlator}
\end{equation}
where ${\cal H}^{s \rightarrow d \gamma}_{eff}$ is given by
Eq.(\ref{eq:eff}).

As usual, our goal is to make a match between the two representations
of the correlation function (\ref{correlator}) at a certain region
$Q_i^2 \sim 1$ GeV$^2$: the operator product expansion (OPE) in powers
of $Q_i^2$ and the phenomenological representation of the dispersion
integrals. The basic idea, supported by ample successful
applications, is that taking into account only the first few terms
in the $Q_i^2$ expansion, complemented by  rather simple assumptions
for the higher mass contributions to the dispersion relation, will
already provide a good estimate to the amplitudes of interest.

The $\Omega^-$ and $\Xi^-$ interpolating fields are given by
\cite{RRY}
\begin{equation}
\eta^{\Xi^-}(x)=-\epsilon_{abc}
\left({s^T_a}(x)C\gamma_\mu s_b(x)\right)
\gamma_5\gamma^\mu d_c(x)\ ,
\end{equation}
\begin{equation}
\eta_\mu^{\Omega^-}(x)=\epsilon_{abc}
\left({s^T_a}(x)C\gamma_\mu s_b(x)\right)s_c(x)\ .
\end{equation}

With these definitions and working at leading order in perturbation
theory we obtain for the T product appearing in Eq.(\ref{correlator})
\begin{eqnarray}
& &
\langle 0|{\rm T}[\eta^\Xi(y){\cal H}^{s \rightarrow d \gamma}_{eff}(0)
\overline{\eta}_\mu^\Omega(x)]
|0\rangle = \frac{-ieG_F}{16\pi^2}0.28m_s\epsilon_{abc}\epsilon_{a'b'c'}
F^{\alpha\beta}\gamma_5\gamma^\nu S^d_{ce}(y)(1+\gamma_5)\sigma_{\alpha
\beta}S^s_{ec'}(-x)\times
\nonumber\\*[7.2pt]
& &
\left\{tr[S^s_{bb'}(y-x)\gamma_\mu C(S^s_{aa'}(y-x))^T
C\gamma_\nu]
+2\gamma_\mu C (S^s_{aa'}(y-x))^T C\gamma_\nu S^s_{bb'}(y-x)\right\} \; ,
\label{T}
\end{eqnarray}
where the coordinate-space quark propagator in the presence of the
quark condensate takes the following form \cite{RRY,HK}
\begin{equation}
S_{ab}^q(x) = \langle 0|{\rm T}[q_a(x)\overline{q}_b(0)]|0\rangle =
{i\delta_{ab}\over 2\pi^2}{\rlap{/}{x}\over
 x^4}-{\delta_{ab}\over 4\pi^2}{m_q\over x^2} -
{\delta_{ab}\over 12}\langle\overline{q}q\rangle +{i\delta_{ab}\over
48}m_q\langle\overline{q}q\rangle\rlap{/}{x}
+ \cdot\cdot\cdot \; .
\label{prop}
\end{equation}

The amplitude $\Pi_\mu(Q_i,Q_f)$ includes a lot of different
non-trivial Lorentz structures. For each one of these structures,
$\Pi_k(Q_i^2,Q_f^2)$, we can write a double dispersion representation
of the form
\begin{equation}
\Pi_k(Q_i^2,Q_f^2) = {1\over\pi^2}\int_0^\infty ds_1\int_0^\infty ds_2
{\rho_k(s_1,s_2)\over (s_1+Q_f^2)(s_2+Q_i^2)} + \cdot\cdot\cdot \; ,
\label{spec}
\end{equation}
where the ellipsis represents subtractions polynomials in $Q_i^2$ and
$Q_f^2$, which will vanish under the double Borel transform \cite{ioffe1},
which is a straightforward generalization of that used in Ref.\cite{SVZ2}.
Applying the double Borel transform to Eq.(\ref{spec}) gives
\begin{equation}
\Pi_k(M_1^2,M_2^2) = {1\over\pi^2}\int_0^\infty ds_1\int_0^\infty ds_2
\rho_k(s_1,s_2)e^{-s_1/M_1^2} e^{-s_2/M_2^2} \; .
\label{specb}
\end{equation}

In the phenomenological side the various Lorentz structures can be
obtained from the consideration of the $\Omega$ and $\Xi$ contribution
to the dispersion sum rule
\begin{equation}
\langle 0|\eta^\Xi|\Xi(Q_f)\rangle\langle\Xi(Q_f)\gamma|
{\cal H}^{s \rightarrow d \gamma}_{eff}|\Omega(Q_i)\rangle\langle\Omega(Q_i)|
\overline{\eta}_\mu^\Omega|0\rangle \; ,
\label{str}
\end{equation}
where the most general, gauge invariant form for the amplitude of the
$\Omega^-\rightarrow\Xi^-\gamma$ decay is given by \cite{Singer}
\begin{eqnarray}
M(\Omega^-\rightarrow\Xi^-\gamma)& =& \langle\Xi(Q_f)\gamma|
{\cal H}^{s \rightarrow d \gamma}_{eff}|\Omega(Q_i)\rangle =
ieG_F\overline{u}^{(\Xi)}(Q_f)[(a_1+
\gamma_5 a_2)(\rlap{/}{q}g^{\mu\nu}-\gamma^\mu q^\nu)
\nonumber\\*[7.2pt]
&+& (b_1+\gamma_5 b_2)(Q_i. qg^{\mu\nu}-Q_i^\mu q^\nu)/M_\Omega]
u_\nu^{(\Omega)}(Q_i)\epsilon_\mu \; ,
\label{M}
\end{eqnarray}
where $q=Q_f-Q_i$, $\epsilon_\mu$ is the polarization of the photon,
and $a_1$, $a_2$, $b_1$ and $b_2$ are the four amplitudes we want to
evaluate. The other matrix elements contained in Eq.(\ref{str}) are
of the form
\begin{equation}
\langle 0|\eta^\Xi|\Xi(Q_f)\rangle = \lambda_\Xi u^{(\Xi)}(Q_f)\; ,
\label{xi}
\end{equation}
\begin{equation}
\langle\Omega(Q_i)|\overline{\eta}_\mu^\Omega|0\rangle =
\lambda_\Omega \overline{u}_\mu^{(\Omega)}(Q_i) \; ,
\label{omega}
\end{equation}
where $u_\mu(p)$ is a Rarita-Schwinger spin-vector satisfying
\begin{equation}
u_\mu^{(\Omega)}(p)\overline{u}_\nu^{(\Omega)}(p)=-\left(g^{\mu\nu}
-{1\over 3}\gamma_\mu\gamma_\nu - {2\over 3M_\Omega^2}p_\mu p_\nu
+ {\gamma_\mu p_\nu - \gamma_\nu p_\mu\over 3M_\Omega}\right)
(\rlap{/}{p}+M_\Omega) \; ,
\label{rarita}
\end{equation}
$u(p)$ is a Dirac spinor and $\lambda_\Xi$, $\lambda_\Omega$ are
the couplings of the currents with the respective hadronic states.

Saturating the correlation function Eq.(\ref{correlator}) with
$\Omega$ and $\Xi$ intermediate states, and using Eqs.(\ref{str}),
(\ref{M}), (\ref{xi}), (\ref{omega}) and (\ref{rarita}) we get
\begin{eqnarray}
& &\Pi_\mu^{(phen)}(Q_i,Q_f) = -ieG_F\lambda_\Xi\lambda_\Omega
{(\rlap{/}{q}+\rlap{/}{Q}_i+M_\Xi)\over Q_f^2+M_\Xi^2}\left[(a_1+\gamma_5 a_2)
(\rlap{/}{q}\epsilon^\beta-\rlap{/}{\epsilon}q^\beta) +
(b_1+\gamma_5 b_2)\times\right.
\nonumber\\*[7.2pt]
& & \left.
(Q_i. q\epsilon^\beta-Q_i.\epsilon q^\beta){1\over M_\Omega}\right]
 \left(g_{\beta\mu}
-{1\over 3}\gamma_\beta\gamma_\mu - {2\over 3M_\Omega^2}Q^i_\beta Q^i_\mu
+ {\gamma_\beta Q^i_\mu - \gamma_\mu Q^i_\beta\over 3M_\Omega}\right)
{(\rlap{/}{Q}_i+M_\Omega)\over Q_i^2+M_\Omega^2} .
\label{ficor}
\end{eqnarray}

In this work we will derive sum rules for the Lorentz structures:
\begin{eqnarray}
a &:&\; \rlap{/}{q}(a_1+\gamma_5 a_2)\sigma_{\alpha\beta}\gamma_\mu
\rlap{/}{Q}_i\epsilon^\alpha q^\beta\; ,
\label{a}\\
b &:&\; \rlap{/}{q}(b_1+\gamma_5 b_2)(Q_i. q \epsilon_\mu - Q_i
.\epsilon q_\mu)\; .
\label{b}
\end{eqnarray}
This choice is based on the fact that the structure a gets perturbative
contribution (which is very important in the case of double dispersion
relation analysed here \cite{ioffe1}) as well as power corrections
contributions, and the structure b is the only one that contributes
solely to the amplitudes $b_1$ and $b_2$. Furthermore, as we will
show, both sum rules are very stable as a function of the Borel masses.

For the continuum contribution we adopt the standard form of
Ref.\cite{ioffe1}, which completes our parametrization of the spectral
density $\rho_k$:
\begin{eqnarray}
{1\over \pi^2}\rho_a(s_1,s_2)& =& {-2\over 3} eG_F\lambda_\Xi
\lambda_\Omega\delta(s_1-M_\Xi^2)\delta(s_2-M_\Omega^2)
\nonumber\\*[7.2pt]
&+&
\theta(s_1-s_\Xi)\theta(s_2-s_\Omega)\rho^{(0)}_a(s_1,s_2)\; ,
\label{rhoa}\\
{1\over \pi^2}\rho_b(s_1,s_2)& =& -ieG_F\lambda_\Xi
\lambda_\Omega\delta(s_1-M_\Xi^2)\delta(s_2-M_\Omega^2)
\nonumber\\*[7.2pt]
& + &
\theta(s_1-s_\Xi)\theta(s_2-s_\Omega)\rho^{(0)}_b(s_1,s_2)\; ,
\label{rhob}
\end{eqnarray}
where $s_\Xi$, $s_\Omega$ are, respectively, the continuum threshold
of the $\Xi$ and $\Omega$ determined in the mass sum rules, and
$\rho_k^{(0)}(s_1,s_2)$ is the free-quark spectral function.

Let us now calculate the left-hand side of the sum rule Eq.(\ref{specb})
using the OPE. We shall consider only the diagrams shown in Fig.2.
The perturbative contribution to the sum rule is shown in Fig.2a
and is obtained by using only the first term in the right-hand side
of Eq.(\ref{prop}) in Eq.(\ref{T}). The resulting expressions for
the two chosen structures are
\begin{eqnarray}
\Pi_\mu^{(a)}(M_1^2,M^2_2)& =& {-3eG_F0.28m_s\over 2^9\pi^6}
{M_1^6M_2^6\over(M_1^2+M_2^2)^3}\rlap{/}{q}(1-\gamma_5)
\sigma_{\alpha\beta}\gamma_\mu\rlap{/}{Q}_i\epsilon^\alpha q^\beta\; ,
\label{sra}\\
\Pi_\mu^{(b)}(M_1^2,M^2_2)& =& {3ieG_F0.28m_s\over 2^8\pi^6}
{M_1^6M_2^8\over(M_1^2+M_2^2)^4}\rlap{/}{q}(1-\gamma_5)
(Q_i. q \epsilon_\mu - Q_i.\epsilon q_\mu)\; .
\label{srb}
\end{eqnarray}

Noticing that Eq.(\ref{specb}) is the double Laplace transformation
in $1/M_i^2$, the free-quark spectral function, $\rho_k^{(0)}(s_1,s_2)$,
can be obtained from Eqs.(\ref{sra}) and (\ref{srb}) by applying
the inverse transformation. It gives
\begin{eqnarray}
{1\over \pi^2}\rho_a^{(0)}(s_1,s_2)& =& {-3eG_F0.28m_s\over 2^{10}
\pi^6}s_1s_2\delta(s_1-s_2) \; ,
\label{rho0a}\\
{1\over \pi^2}\rho_b(s_1,s_2)& =& {-ieG_F0.28m_s\over 2^9\pi^6}
s_2^3\delta'(s_2-s_1) \; .
\label{rho0b}
\end{eqnarray}

The next term in the OPE is determined by Figs.2b, c and d. The
contribution of Fig.2d exactly cancells the contributions of
Figs.2b and c for the structure b. The result of the three diagrams
for the structure a is
\begin{equation}
{5eG_F0.28m_s\over 2^6\pi^4}m_s\langle\overline s s\rangle
{M_1^2M_2^2\over(M_1^2+M_2^2)^2}\left[M_2^2+M_1^2\ln{\left({
M_1^2+M_2^2\over m_s^2}\right)}\right]\rlap{/}{q}(1-\gamma_5)
\sigma_{\alpha\beta}\gamma_\mu\rlap{/}{Q}_i\epsilon^\alpha q^\beta\; .
\end{equation}

The last diagram which we take into account is shown in Fig.2e. It does
not give any contribution to the structure b and the contribution
to the structure a is
\begin{equation}
{-eG_F0.28m_s\over 12\pi^2}\langle\overline s s\rangle^2
\rlap{/}{q}(1-\gamma_5)
\sigma_{\alpha\beta}\gamma_\mu\rlap{/}{Q}_i\epsilon^\alpha q^\beta\; .
\end{equation}
The gluon condensate is not taken into account because its contribution
is suppressed relative to the operator $m_q\langle\overline q q\rangle$
(which has the same dimension) by an extra loop factor $1/16\pi^2$.

In general, when a real photon is emitted, one has also to consider
diagrams such as that in Figure 3, involving long distances in the photon
channel. However, this diagram gives a negligible contribution, as it
is proportional to the part of the photon wave function at long
distances, involving quarks of different flavors \cite{russos,BBK}.
Therefore, we do not consider it here.

Collecting all the obtained contributions, and transferring the
continuum contribution to the OPE side, we arrive at the following
representation to the amplitudes:
\begin{eqnarray}
a_1=-a_2&=&{3\over2\pi^2}{e^{M_\Xi^2/M_1^2}e^{M_\Omega^2/M_2^2}\over
\tilde\lambda_\Xi\tilde\lambda_\Omega}0.28m_s\left[{3\over2^6}
\int_0^{s_\Xi} ds_1\int_0^{s_\Omega} ds_2 s_1 s_2\delta(s_1-s_2)
e^{-s_1/M_1^2}e^{-s_2/M_2^2}\right.
\nonumber\\*[7.2pt]
&+&\left.{5\pi^2\over16}m_s f a
{M_1^2M_2^2\over(M_1^2+M_2^2)^2}\left(M_2^2+M_1^2\ln{\left({
M_1^2+M_2^2\over m_s^2}\right)}\right) + {1\over 12}f^2a^2\right]\; ,
\label{a1}
\end{eqnarray}
\begin{equation}
b_1=-b_2={1\over2^5\pi^2}{e^{M_\Xi^2/M_1^2}e^{M_\Omega^2/M_2^2}\over
\tilde\lambda_\Xi\tilde\lambda_\Omega}0.28m_s
\int_0^{s_\Xi} ds_1\int_0^{s_\Omega} ds_2 s_2^3{d\over ds_2}(
\delta(s_2-s_1)) e^{-s_1/M_1^2}e^{-s_2/M_2^2} \; ,
\label{b1}
\end{equation}
where $\tilde\lambda_H=4\pi^2\lambda_H$, $a=-4\pi^2
\langle\overline q q\rangle$, and $f=\langle\overline s s\rangle/
\langle\overline q q\rangle$.

\subsection{Sum Rule Analysis}
The sum rules are sampled in the region of Borel mass, $M^2$,
which have been identified as the fiducial region for the
baryon mass sum rules \cite{ioffe2}
\begin{eqnarray}
& &
1.4\leq M_1^2\leq 2.0\, {\mbox{GeV}}^2\hspace*{2cm}
{\mbox{for}}\, \Xi \, ,
\\*[7.2pt]
& &
2.4\leq M_2^2\leq 3.2\, {\mbox{GeV}}^2\hspace*{2cm}
{\mbox{for}}\,\Omega \, .
\end{eqnarray}
The values used for $\tilde\lambda_H$ and $s_H$ are extracted
from the respective mass sum rules analysed in the same region
given above. For $\Xi$ we get from Ref.\cite{ioffe2}
$\tilde\lambda_\Xi\simeq 1.6{\mbox{GeV}}^3$ and $s_\Xi=
3.6{\mbox{GeV}}^2$. For $\Omega$, using the sum rules given in
\cite{espriu}, we get $\tilde\lambda_\Omega\simeq 3.8{\mbox{GeV}}^3$
and $s_\Omega\simeq 7.5{\mbox{GeV}}^2$. The value of the other
parameters are \cite{ioffe2}: $a=0.55\, {\mbox{GeV}}^3 ,\;
m_s=150 {\mbox{MeV}} ,\; f=0.8 ,\; M_\Xi=1.32 {\mbox{GeV}}$
and $M_\Omega=1.67 {\mbox{GeV}}$.

In Fig.4 we show the result obtained for $a_1$ as a function of
$M_1^2$ for different values of $M_2^2$. One can see that $a_1$
varies very slowly with $M_1^2$ but not so slowly with $M_2^2$. In the
Borel region considered the amplitude $a_1$ change less than $20\%$.
The same behaviour can be observed in Fig.5 where $a_1$ is plotted
as a function of $M_2^2$ for different values of $M_1^2$.

For the amplitude $b_1$ we get an even more stable result. As
can be seen by Figs.6 and 7 $b_1$ varies slowly with $M_2^2$ and
with $M_1^2$. In the Borel region considered $b_1$ changes less
than $10\%$. Using $M_1^2=M_\Xi^2$ and $M_2^2=M_\Omega^2$ we
get
\begin{eqnarray}
& &
a_1 = -a_2 \simeq 1.10 {\mbox{MeV}} \, ,
\\*[7.2pt]
& &
b_1 = -b_2 \simeq -0.58 {\mbox{MeV}} \, .
\end{eqnarray}
\section{Results and Conclusions}
Given the values of the form factors $a_1,a_2,b_1,b_2$, we can compute the
decay width $\Gamma (\Omega^- \rightarrow \Xi^- \gamma)$ \cite{Singer}:
\begin{equation}
\Gamma (\Omega^- \rightarrow \Xi^- \gamma) = \frac{2}{3} \alpha G_F^2 q_0^3 H
\end{equation}
where
\begin{equation}
q_0 = \frac{(M_{\Omega}^2 - M_{\Xi}^2)}{2 M_{\Omega}}
\end{equation}
is the photon energy in the $\Omega$ rest frame and
\begin{eqnarray}
H &=& \left[ (1 + \frac{p_{\Omega} \cdot p_{\Xi}}{M_{\Omega}^2} )
(a_1^2 + a_2^2) +
( \frac{p_{\Omega} \cdot p_{\Xi}}{M_{\Omega}^2} ) (b_1^2 + b_2^2) +
\right. \\ \nonumber
& & \left. (1 +  \frac{p_{\Omega} \cdot p_{\Xi}}{M_{\Omega}^2})
(a_1 b_1 + a_2 b_2) +
\frac{M_{\Xi}}{M_{\Omega}} (b_1^2 - b_2^2) +
\frac{2 M_{\Xi}}{M_{\Omega}} (a_1 b_1 - a_2 b_2)  \right]
\end{eqnarray}
Using the values obtained in the last section we get:
\begin{equation}
\Gamma (\Omega^- \rightarrow \Xi^- \gamma) = 5.6 \times 10^{-11} \mbox{eV}
\end{equation}
which results in a branching ratio
\begin{equation}
\mbox{BR} (\Omega^- \rightarrow \Xi^- \gamma) = 7.0 \times 10^{-6} .
\label{result}
\end{equation}

As mentioned before, the penguin contributions to this decay were
calculated using bag model matrix
elements, for weak and electromagnetic transitions at baryon level,
and standard penguin coefficients, yielding \cite{KE}:
\begin{equation}
\mbox{BR}^{Penguin} (\Omega^- \rightarrow \Xi^- \gamma) \sim 2
\times 10^{-3}(C_P/C_1)^2 \simeq 5 \times 10^{-6} ,
\label{eeg}
\end{equation}
where the ratio of the $C_P/C_1$ coefficients of the QCD-corrected
nonleptonic Hamiltonian is approximately 1/20 \cite{SVZ2,gw,gp}.

Our result suggests that the different contributions to the branching ratio
$BR(\Omega^- \rightarrow \Xi^- \gamma)$ arising from the single-quark
 $s\rightarrow d\gamma$ transition, Eq.(\ref{result}), the penguin diagram,
Eq.(\ref{eeg}), and the long distance
process, Eq.(\ref{long}), are all comparable. In order
to separate the contributions from these different processes it may
be necessary to study the asymmetry parameters of the decay. It is also
worth mentioning that the result given in Ref.\cite{ara}, Eq.(\ref{arasin}),
would be of the same order as our result, Eq.(\ref{result}), had they
used the same value of $c_7(m_s)$ given here. In Ref.\cite{ara} they
also estimated another class of long distance contributions to the $\Omega^-
\rightarrow \Xi^- \gamma$ decay, using the vector meson dominance
approximation,
and they found that it could possibly saturate the experimental upper bound.
However, their long distance result is proportional to a quantum
chromodynamics coefficient, whose direct estimate is not reliable since
all the calculation is far from the perturbative regime. For this reason
we chose to quote here only the more traditional long distance result
given in Ref.{\cite{KS}.

Our result is roughly a factor
of $60$ below the experimental upper limit, Eq.(\ref{exp}),
and it would be extremely
interesting if new experiments could bring the upper limit down, since
even adding all the contributions to this decay, one still gets a number
that is at least one order of magnitude smaller than the experimental
upper limit.
\section{Acknowledgments}
We would like to thank I.F. Albuquerque for useful discussions
and for insisting on the relevance of the subject. R.R. would like to thank
A. Ioannissian for useful correspondence.
This work was partially supported by CNPq and FAPESP.

%

%
%
\begin{table}
\caption{Values of $F_2$ and $ | \lambda_i | F_2$ for different internal
quarks. }
\end{table}

\begin{figure}
\caption{Typical diagrams contributing to hyperon radiative decays
at quark level.}
\label{fig-1}
\end{figure}
\begin{figure}
\caption{Diagrams considered for the calculation of the Wilson
coefficients of the correlator function. The crossed circle indicates
the weak transition, with photon emission, and the cross on the s-quark
line indicates that the $m_s$ correction to the propagator is
relevant.}
\label{fig-2}
\end{figure}
\begin{figure}
\caption{A $q\overline{q}$ pair in the vacuum coupled to the
electromagnetic current at large distances.}
\label{fig-3}
\end{figure}
\begin{figure}
\caption{The amplitude $a_1$ as a function of $M_1^2$ for $M_2^2=
M_\Omega^2$ (full line),$M_2^2= 2.4$GeV$^2$ (dashed line) and
$M_2^2=3.2$GeV$^2$ (dotted line).}
\label{fig-4}
\end{figure}
\begin{figure}
\caption{The amplitude $a_1$ as a function of $M_2^2$ for $M_1^2=
M_\Xi^2$ (full line),$M_2^2= 1.4$GeV$^2$ (dashed line) and
$M_2^2=2.0$GeV$^2$ (dotted line).}
\label{fig-5}
\end{figure}
\begin{figure}
\caption{Same as Figure 4 for $b_1$.}
\label{fig-6}
\end{figure}
\begin{figure}
\caption{Same as Figure 5 for $b_1$.}
\label{fig-7}
\end{figure}
\newpage

\begin{center}

\begin{tabular}{|c|c|c|}  \hline
Quark      &  $F_2$      & $| \lambda_i | F_2$ \\  \hline  \hline
u          &  $2.3 \times 10^{-9}  $  &  $ 4.8 \times 10^{-10}$ \\   \hline
c          &  $2.0 \times 10^{-4}  $  &  $ 4.2 \times 10^{-5}$ \\  \hline
t          &  $0.39$                  &  $ (4.7 - 28 ) \times 10^{-5}$ \\
\hline
\end{tabular}

\end{center}

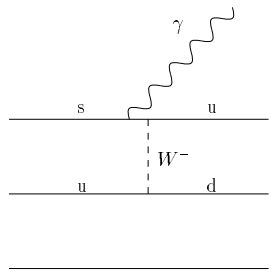
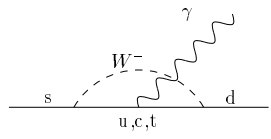
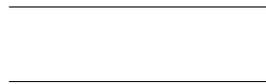
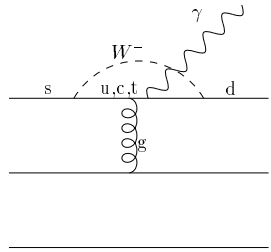
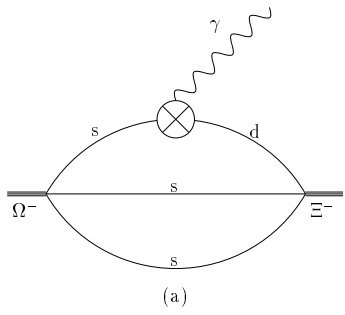
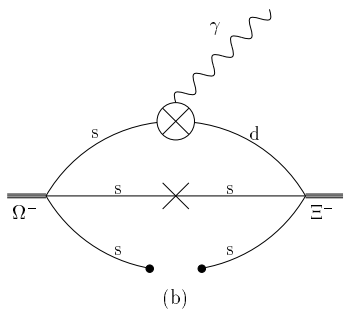
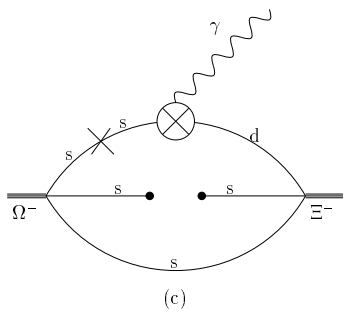
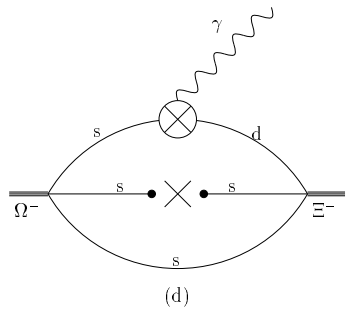
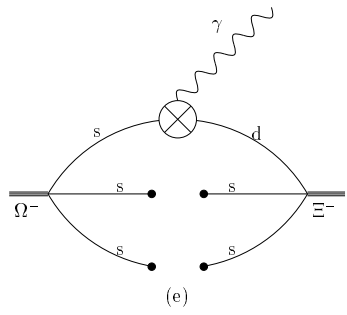
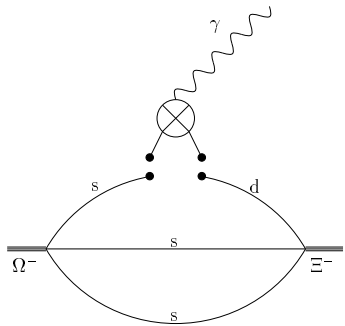
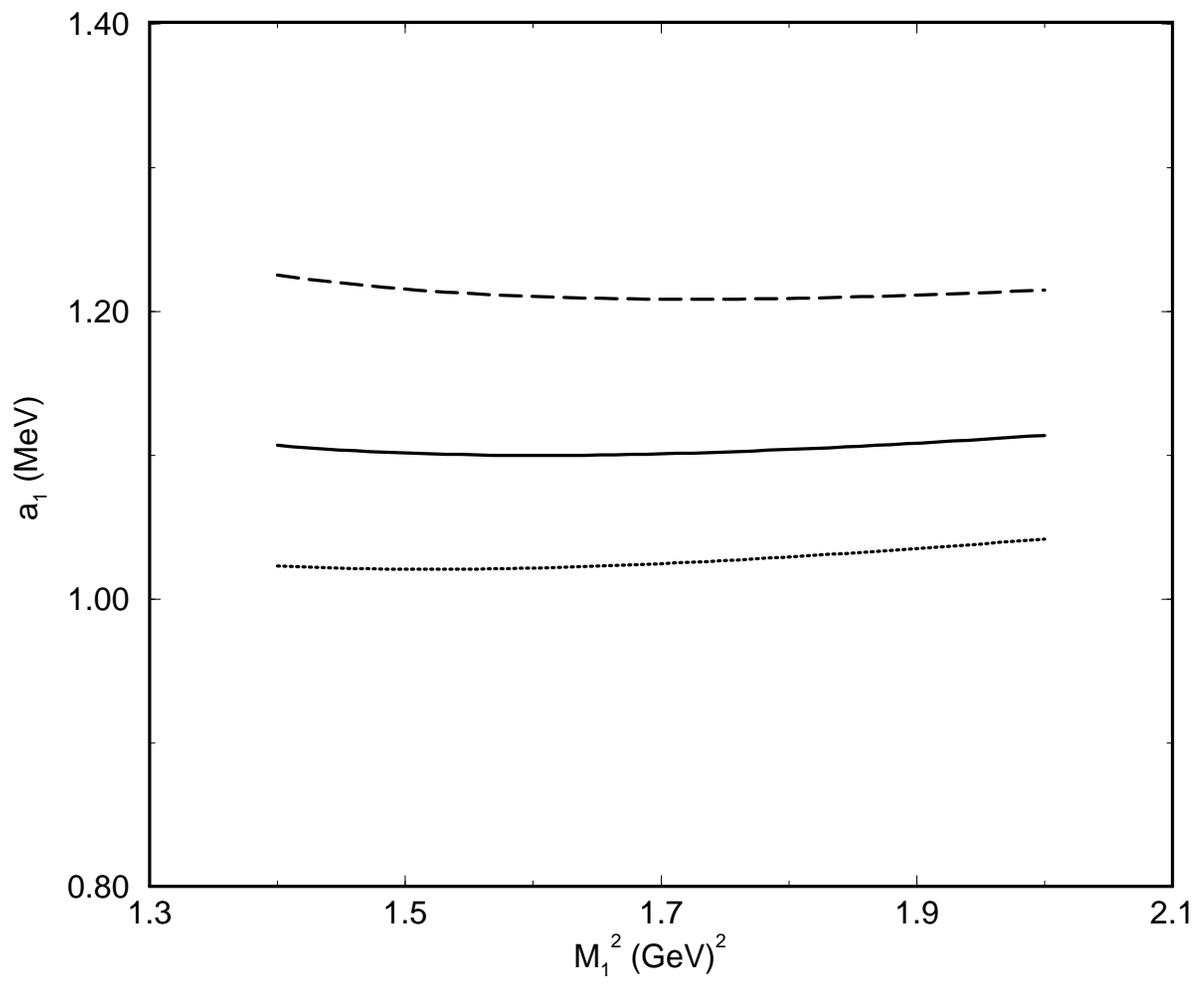
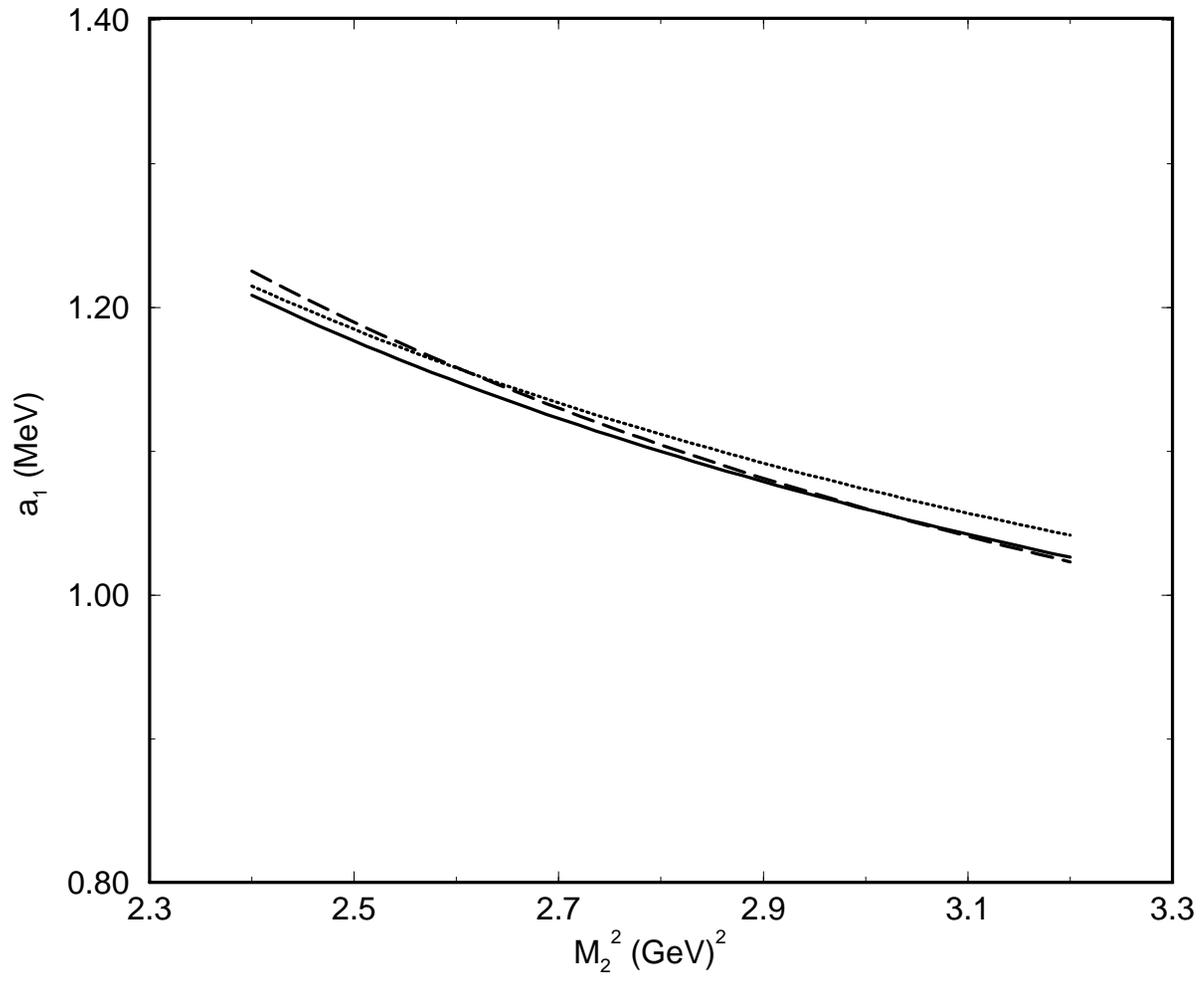
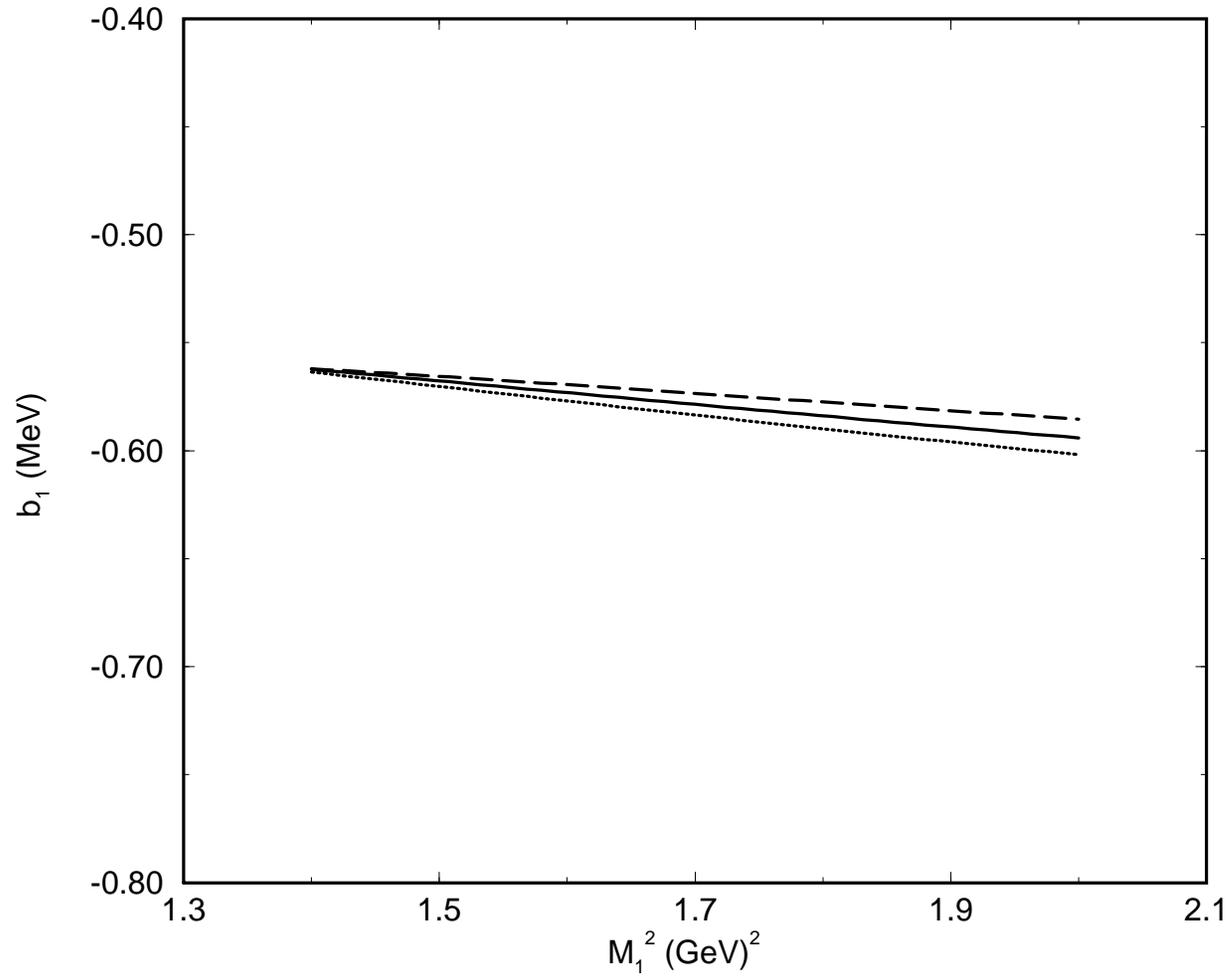
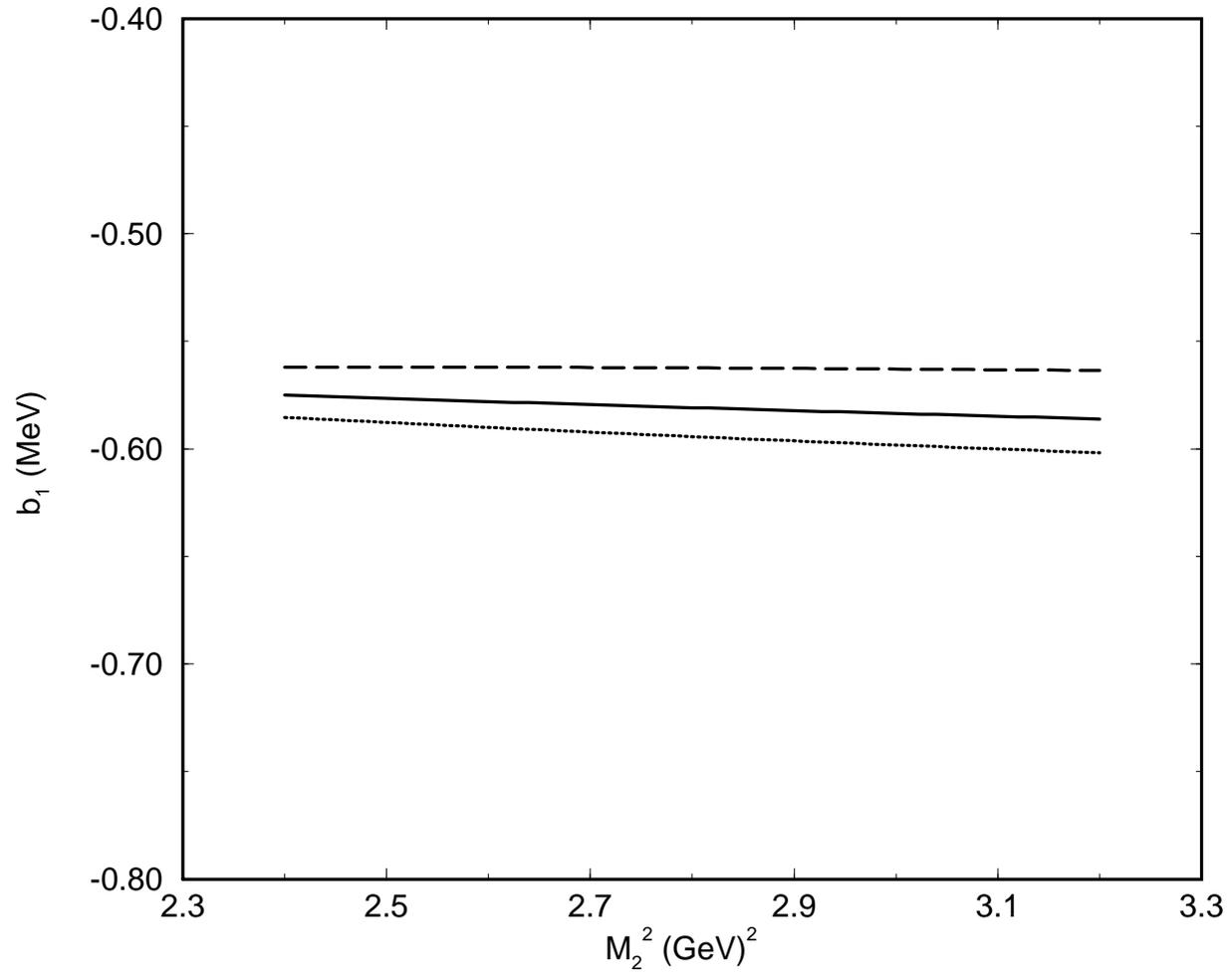
\end{document}